\documentclass[aps,prd,twocolumn,groupedaddress]{revtex4}

\usepackage{graphicx}
\usepackage{dcolumn}
\usepackage{bm}

\begin{document}

\preprint{}

\title{Dark Energy as an Inverse Problem}

\author{Cristina Espa\~{n}a-Bonet}
\email[]{cespana@am.ub.es}
\affiliation{Department of Astronomy, University of Barcelona
\\ and CER for Astrophysics, Particle Physics and Cosmology, 
Mart\ii~i Franqu\`es 1, Barcelona 08028, Spain}

\author{Pilar Ruiz-Lapuente}
\email[]{pilar@am.ub.es}
\affiliation{Department of Astronomy, University of Barcelona
\\ and CER for Astrophysics, Particle Physics and Cosmology, 
Mart\ii~i Franqu\`es 1, Barcelona 08028, Spain}
\affiliation{Max-Planck-Institut f\"ur Astrophysik, Karl Schwarzschild 
Stra$\beta$e 1, 85740 Garching, Germany}

\date{\today}

\newcommand{\ii}{\'\i}
\newcommand{\M}[1]{\textrm{\textbf{#1}}}

\begin{abstract}
A model--independent approach to dark energy is here developed by 
considering the determination of its equation of state 
as an inverse problem. The reconstruction of $w(z)$ as a non--parametric 
function using the current SNe Ia data is explored. It is investigated
as well how results would improve when considering other samples of 
cosmic distance indicators at higher redshift. This approach reveals
the lack of information in the present samples to conclude on the
behavior of $w(z)$ at z $>$ 0.6. At low level of significance a
preference is found for $w_{0}$ $<$ --1 and  $w'(z)$ $>$ 0 at z
$\sim$ 0.2--0.3. The solution of $w(z)$ along redshift never departs
more than 1.95 $\sigma$ from the cosmological constant $w(z)=-1$, and
this only occurs when using various cosmic distance indicators.  
The determination of $w(z)$ as a function is readdressed
considering samples of large number of SNe Ia as those to be provided 
by SNAP. It is found an improvement 
in the resolution of $w(z)$ when using those synthetic samples,
 which is favored by adding data at very high z. The
set of degenerate solutions compatible with the data can be retrieved 
through this method.

\end{abstract}


\maketitle

\section{Introduction}
\label{intro}

The acceleration of the rate of expansion of the Universe,  
first discovered through supernovae \cite{Riess98,p99}
is being confirmed by  new cosmological tests. The cosmic microwave
background measurements by the Wilkinson Microwave Anisotropic Probe
(WMAP) \cite{wmap} and results from the large scale 
distribution of galaxies \cite{lss} and galaxy clusters \cite{Xcumuls}
confirm that our universe is dominated by negative pressure. 
The understanding of the accelerated expansion of the cosmos might result
in a change of the framework in
which gravity is to be described or in the identification 
of unknown components of the cosmos. 
All this seems most fundamental to physics and cosmology, thus 
large samples of cosmological data and  various procedures
to analyse them are being examined. 

In a descriptive way, the component or new physics responsible for
 the acceleration of the expansion, the so called
``dark energy'', can be incorporated in the right--hand side of the 
 Friedmann--Robertson--Walker (FRW) equations and
be simply addressed as an additional term for whom we intend to 
determine the barotropic index: $w(z)=p(z)/\rho(z)$.

The reconstruction of $w(z)$ from a given sample of data has been attempted
proposing fitting functions or expansion series of $w(z)$ along z 
in ways to accomodate a wide
range of dark energy candidates 
\cite{saini0,gerke,Linder,cdist2,visser,Nesseris,Huterer2,Huterer3,alam1,alam2,
jonsson,bassett1,parkinson,bassett2,feng}. 
There has been some debate on the effect  that choosing particular 
models for those functions or truncating the expansion series in z
might have in deriving possible evolution \cite{jonsson,Linder2,alam1,
alam2,alam3}.

Here, we have developed an approach to obtain $w(z)$ without imposing
any constraints on the form of the function. This is addressed through
a generalized nonlinear inverse approach. This method allows to examine
the resolution of the equation of state $w(z)$ at 
various redshifts and 
through various samples. One can quantify the improvement in information
provided by increasing a sample or by the addition of various sources.
The inverse approach formulated by Backus and
Gilbert(1970)\cite{backus} 
has been 
widely used in geophysics and solar structure physics. 
In this approach,  
the mere fact that  the continuous functional has to be derived from a 
discrete number of data implies a non--uniqueness of the answer.
It has also been  shown that, 
even if the data were dense and with
no uncertainty, there would be more than one solution to many specific
inverse problems such as the determination of the density structure of 
the earth from the data on the local gravitational field  
at its surface, and others. 
This lack of uniqueness comes from the way in which the different
equations reflect in the observables used. The problem 
of the determination of dark energy faces such degeneracy. 
In the luminosity distance along z from supernovae  
and other cosmic distance indicators, $w(z)$ enters in an  
integral form, which limits the possibility to access to $w(z)$. 

In earlier examinations of the degeneracy in $w(z)$ obtained through
cosmic distance indicators a range of solutions giving the same 
luminosity distance along z were pointed out \cite{maor1, maor2}. 
As more data would constrain $w(z)$ at various redshifts, not
only using the distance luminosity, but other indicators as well, the
reconstruction should become more successful.

To compare with a significant body of work which analyses the data 
using the expansion to first order  $w(z)= w_0 + w_a z/(1+z)$, we will
also formulate  this approach for the case of determination of
discrete parameters. This allows to quantify the increase of
information among SNe Ia samples and samples of other distance indicators.
We examine from current data 
the possibility of determining at present the values of $w(z)$ 
and its first derivative and compare with previous results
\cite{jonsson,Riess04}.

\section{Inverse problem}
\label{sec_IP}

\subsection{Non--parametric non--linear inversion}
\label{continuous}

The inverse problem provides a powerful way to determine the values
of functional forms from a set of observables. This approach is useful
when the information along a certain coordinate, in our case
information on $w(z)$, emerges in observables
coupled with information at all other z. We have a surface picture
of $w(z)$ in the luminosity distance at a given z, 
as the p--mode waves in the Sun surface have on 
its internal structure. Dark energy is here addressed using the
non--linear non--parametric inversion. Most frequently, 
when the parameters to be determined
are a set of discrete unknowns, the method used is a least
squares. But the continuos case, where functional forms are to be 
determined, requires a general inverse problem formulation.
The inverse method used here is a Bayesian 
approach to this generalization \cite{tarantola}.

We consider a flat universe with only two dominant constituents
(at present): cold matter and dark energy. Therefore we characterize 
the cosmological model by the density of matter, $\Omega_M$, and 
by the index $w(z)$ of the dark energy equation of state,
\begin{equation} 
w(z)=\frac{p(z)}{\rho(z)}
\end{equation}
The vector of unknowns $\M{M}$ has then a discrete and a continuous 
component,
\begin{equation}
      \M{M}  = \left( \begin{array}{c}
                  \Omega_M        \\
                  w(z)            \\
                  \end{array}
           \right)\
\end{equation}

\noindent
On the other hand,
the observational data are mainly SNe\,Ia magnitudes. 
We have a finite set of $N$ magnitudes, $m_i$, and consider the following 
theoretical equation, the magnitude-redshift relation in a flat
universe relating the unknowns to the observational data: 

\begin{equation}
m^{th}(z,\Omega_M,w(z)) = {\cal M} ~ +
  5\log[D_L(z,\Omega_M,w(z))], 
\label{mag}
\end{equation}
\noindent where 
\begin{equation}
{\cal M} \equiv M -5\log H_0 +25
\label{calmag}
\end{equation}
\noindent and $D_L$ is the Hubble-free luminosity distance
\begin{equation}
D_L(z,\Omega_M,w(z)) = c(1+z) 
\int_{0}^{z}{\frac{H_0 dz^\prime}{H(z^\prime,\Omega_{M},w(z))}}
\label{dl}
\end{equation}
with

\begin{eqnarray}
H(z^\prime,\Omega_{M},w(z)) =
H_0~\sqrt{\Omega_{M}(1+z^\prime)^3 + \Omega_{X}(z^\prime)} 
\label{hubblep}
\end{eqnarray}

\begin{equation}
\Omega_{X}(z)=\Omega_{X}\exp{\left( 3 \int_{0}^{z}
{dz^{\prime}\frac{1+w(z^{\prime})}{1+z^{\prime}} }\right)}.
\label{densol}
\end{equation}

In order to combine the results with
other data we substitute the original SNe magnitudes by the
dimensionless distance coordinate $y$: 

\begin{equation}
y_i\equiv\frac{exp_{10}{((m_i-{\cal M})/5)}}{c(1+z_i)} =
\int_{0}^{z_i}{\frac{dz^\prime}{\sqrt{\Omega_{M}(1+z^\prime)^3 +
\Omega_{X}(z^\prime) }}},
\label{f}
\end{equation}

\begin{equation}
\sigma_{y_i} = \frac{\ln 10}{5} y_i (\sigma_{m_i}+\sigma_{\cal M}).
\label{sf}
\end{equation}

\noindent
With this definition we deal directly with a function $y(\Omega_M,w(z))$, 
the only part which depends on the cosmological model. 
To convert our $m_i$ data to $y_i$, 
we can adopt the value obtained from
low redshift supernovae and use ${\cal M}=-3.40\pm0.05$.
Defined in this way, $y_i$ is used in other analyses \cite{cdist2}.

\noindent
After the corresponding transformations, the observables
 are now described by a 
vector of $N$ components, $y_i$, and by a covariance matrix
 ($\M{C}_y$).
  This method can handle
correlated measurements, where non--diagonal elements 
$ C_{y_{i}y_{j}}$ are different from zero
(observations i and j being correlated). But, at present, those have not been
estimated for the composite samples of distance indicators. We would
then use: 

\begin{equation}
\label{coy}
      C_{y,ij} =  \sigma^2_{y_{ij}}  \delta_{ij}
\end{equation}

\noindent
The unknown vector of   
parameters is described by its {\it a priori} value, $\M{M}_0$, 
and the covariance matrix ($\M{C}_0$). The function describing $w(z)$
should be smooth. This leads to no null covariance between neighboring
points in z for $w(z)$ (the smoothness of $w(z)$ implies that if at z,
the value of $w(z)$ has a deviation $w(z) - w(z')$ of a given sign
and magnitude, we want, at a neighboring point z$'$, the deviation 
$w(z) - w(z')$ to have a similar sign and magnitude.

Thus, the covariance matrix $\M{C}_0$ has the form:

\begin{equation}
\label{com}
      \M{C}_0 = \left( \begin{array}{cc}
                     \sigma^2_{\Omega_M}   & 0        \\
                     0            	   & C_{w(z),w(z')} \\
            \end{array} \right)
\end{equation}

\noindent
where a  choice is made for the non--null covariance between  z and z$'$,
 $C_{w(z),w(z')}$. This choice is taken to be as general as possible.
It would define the smoothness required in the solution by setting the
correlation length between errors in z and z$'$ (this gives the length
scale in which the function can fluctuate between redshifts). The
 amplitude of the 
fluctuation of the function is given by the dispersion $\sigma_w$ at z.
In the Gaussian choice for  $C_{w(z),w(z')}$, 
$\sigma_w$ is the 1$\sigma$ region where the solution is to be found.

Thus for a Gaussian 
choice,  $C_{w(z),w(z')}$  is described as:

\begin{equation}
C_{w(z),w(z')}=\sigma_w^2\exp\left(-\frac{(z-z')^2}{2\Delta_z^2}\right),
\label{covwgauss} 
\end{equation} 

\noindent
which means that the variance at z equals $\sigma_w^2$ and that the 
correlation length between errors is $\Delta_z$. 
Another possible choice for $C_{w(z),w(z')}$
is an exponential of the type: 

\begin{equation}
C_{w(z),w(z')}=\sigma_w^2\exp\left(-\frac{|z-z'|}{\Delta_z}\right),
\label{covwexp} 
\end{equation} 

\noindent
while no difference in the results is found by those different choices
of $C_{w(z),w(z')}$.

We are interested in determining the best estimator ${\tilde{\M{M}}}$ 
for \M{M}.
The probabilistic approach incorporates constraints from 
priors through the Bayes theorem, i.e, the {\it a posteriori} 
probability density  $f_{post}(\M{M}/\M{D})$
for the vector \M{M} containing the unknown model parameters given the
observed data \M{D}, is linked to the likelihood 
function L and the prior density function for the parameter vector as: 

\begin{eqnarray}
 f_{post}(\M{M}/\M{D}) \, \, \alpha \, {\it L}(\M{D}/\M{M}) \, \, 
\cdot \, f_{prior}(\M{M}) 
\label{chi0}
\end{eqnarray} 

The theoretical model described by the operator $\M{y}^{th}$, which 
connects the model parameters \M{M} with the predicted data
$ \M{D}_{predicted} = \M{y}^{th}(\M{M})$ is to agree as closely as possible 
with the observed data $\M{y}$. Assuming that both the prior probability and 
the errors in the data are distributed as Gaussian functions, the
posterior distribution becomes:

\begin{eqnarray}
 f_{post}(\M{M}/\M{y}) \, \alpha \, exp[- 
\frac{1}{2}~(\M{y}-\M{y}^{th}(\M{M}))^* ~ \M{C}_y^{-1} ~ 
(\M{y}-\M{y}^{th}(\M{M})) \nonumber   \\
- \frac{1}{2}~(\M{M}-\M{M}_0)^* ~ \M{C}_0^{-1} ~ (\M{M}-\M{M}_0)] \nonumber \\ 
\label{chi5}
\end{eqnarray} 

\noindent
where $^{*}$ stands for the adjoint operator. 
The best estimator for \M{M}, $\tilde{\M{M}}$, is the most probable value of
\M{M}, given the set of data $\M{y}$. The condition is reached by 
minimizing the misfit function:

\begin{eqnarray}
S \equiv \frac{1}{2}~(\M{y}-\M{y}^{th}(\M{M}))^* ~ \M{C}_y^{-1} ~ 
(\M{y}-\M{y}^{th}(\M{M})) +  \nonumber \\
\frac{1}{2}~(\M{M}-\M{M}_0)^* ~ \M{C}_0^{-1} ~ (\M{M}-\M{M}_0),
\label{chi}
\end{eqnarray} 

\noindent
which is equivalent to maximize the Gaussian density of 
probability when data and parameters are treated in the same way. 
This Bayesian approach helps to regularize the
inversion.

Let us now define the operator $\M{G}$ represented by the matrix of partial 
derivatives of the dimensionless distance coordinate, which will simplify 
subsequent notation. Its kernel will be denoted by $g$.

\begin{equation}
\label{g}
\M{G} = \left( \begin{array}{cc}
                    \frac{\partial y_{1}^{th}}{\partial \Omega_M} & 
			\frac{\partial y_{1}^{th}}{\partial w(z)}      \\
                     \frac{\partial y_{2}^{th}}{\partial \Omega_M} & 
			\frac{\partial y_{2}^{th}}{\partial w(z)}     \\
			: & :	\\
		     \frac{\partial y_{N}^{th}}{\partial \Omega_M} &
			\frac{\partial y_{N}^{th}}{\partial w(z)}
                 \end{array}
           \right) 
\end{equation}

\noindent
with

\begin{eqnarray}
 \frac{\partial y_{i}^{th}}{\partial \Omega_M}&=& -\frac{1}{2}
\int_{0}^{z_i}\frac{(1+z')^3 dz'}{H^3(z')} \nonumber \\
& \equiv & \int_{0}^{z_{i}} g_{\Omega_M}(z^\prime)dz^\prime,
\label{derom}
\end{eqnarray}

\begin{eqnarray}
\frac{\partial y_{i}^{th}}{\partial w(z)}&=& -\frac{1}{2}
\int_{0}^{z_i}\frac{3\Omega_X(z')\ln(1+z') dz'}{H^3(z')} \nonumber \\
&\equiv & \int_{0}^{z_{i}} g_{w}(z^\prime)dz^\prime.
\label{derwz}
\end{eqnarray}

\noindent
As shown in Eq.\,\ref{mag} or equivalently Eq.\,\ref{f}, the inverse 
problem is nonlinear in the parameters, thus the solution 
is reached iteratively in a gradient based search. To minimize 
$S$ in Eq.\,\ref{chi}, one demands stationarity. For the non--linear 
case the solution has to be implemented as an iterative procedure where
 \cite{tarantola}: 

\begin{eqnarray}
\label{solutionMit}
\tilde{\M{M}}_{[k+1]} = \M{M}_0 ~ + ~ \M{C}_0 ~ \M{G}^*_{[k]} ~ 
(\M{C}_y + \M{G}_{[k]} \M{C}_0 ~ \M{G}^*_{[k]})^{-1}
 \nonumber \\
(\,\M{y} ~ - ~ \M{y}^{th}(\tilde{\M{M}}_{[k]}) ~+~ \M{G}_{[k]} ~ 
(\tilde{\M{M}}_{[k]}-\M{M}_0)\,)
\end{eqnarray}

\noindent
The estimate of the dimensionless
distance coordinates in the iterations is given by:

\begin{eqnarray}
\label{solutionY}
\tilde{\M{y}}_{[k+1]} = \M{y} ~ - ~ \M{C}_y ~ 
(\M{C}_y + \M{G}_{[k]} \M{C}_0 ~ \M{G}^*_{[k]})^{-1}
 \nonumber \\
(\,\M{y} ~ - ~ \M{y}^{th}(\tilde{\M{M}}_{[k]}) ~+~ \M{G}_{[k]} ~ 
(\tilde{\M{M}}_{[k]}-\M{M}_0)\,)
\end{eqnarray}

\noindent
Since we are
working in a Hilbert space with vectors containing functional forms,
the above operator products give rise to scalar 
products of the functions integrated over the domain of those 
functions. The expressions transform into having the above products
rewritten in terms of the kernels of the operators
\cite{nercessian}.

\noindent
We will indicate the scalar product by `` $\cdot$ '' and it is 
defined as it can be seen from this example:

\begin{equation}
 C_w \cdot \frac{\partial y_{j}^{th}}{\partial w(z)} = 
\int_{0}^{z_j}dz^\prime C_w(z,z^\prime) g_w(z^\prime),
\end{equation}

The components of
the vector of unknowns $\tilde{\M{M}}$, which in our case will be 
both $\Omega_M$ and $w(z)$, are then obtained from:

\begin{eqnarray}
& &\tilde{M}_{[k+1]}(z) = M_0(z) + \nonumber \\
& &\sum_{i=1}^N W_{i[k]} \int_{0}^{z_i}C_{0}(z,z')g_{i[k]}(z')dz'
\end{eqnarray}
where 

\begin{equation}
\label{w}
W_{i[k]} =  \sum_{j=1}^N \left(S_{[k]}^{-1}\right)_{i,j} V_{j[k]} 
\end{equation}

\begin{eqnarray}
\label{Sgen2}
  \M{V}~~~ & = & \M{y} + ~\M{G}~ (\M{M}-\M{M}_0) - \M{y}^{th}(\M{M}) \nonumber \\
  V_{i[k]} & = & y_{i} + \int_{0}^{z_i} g_{j[k]}(z) \left( M_{[k]}(z)-M_0(z) \right)dz \nonumber \\
           &  & - \, y_{i}^{th}(z_i,\Omega_M,w(z)) \\
\nonumber \\
  \M{S}~~~~~\, & = & \M{C}_y ~ + ~ \M{G}~ \M{C}_0 ~\M{G}^*  \nonumber \\
  S_{i,j[k]} & = & (C_y)_{i,j} + \nonumber \\
             &  & \int_{0}^{z_j}\int_{0}^{z_i} g_{i[k]}(z) \,C_0(z,z')\, g_{j[k]}(z') \,dz\, dz' \nonumber \\
  & &
\end{eqnarray} 

In the case of the dark energy equation of state and the matter
density the expressions reduce to 

\begin{equation}
\Omega_{M[k+1]}= \Omega_{Mo} + \sigma_{\Omega_M}^2 \sum_{i=1}^N W_{i\,[k]}\,  
{\frac{\partial y_i^{th}}{\partial \Omega_M}}_{[k]}
\label{om}
\end{equation}

\begin{equation}
w_{[k+1]}(z) = w_o(z)+ \sum_{i=1}^N W_{i\,[k]} 
              \int_{0}^{z_i}C_{w}(z,z')g_w(z')_{[k]}dz'
\label{wz}
\end{equation}

\noindent
where
 $C_{w}(z,z')\equiv C_{w(z),w(z')}(z,z')$, $ W_{i\,[k]}$
is given by the product (\ref{w}) with:


{\setlength\arraycolsep{2pt}  
\begin{eqnarray}
V_i & = & y_{i} + \frac{\partial y_{i}^{th}}{\partial \Omega_M} 
(\Omega_M-\Omega_{M_0}) + \frac{\partial y_{i}^{th}}{\partial w(z)} \cdot 
(w-w_o)  \nonumber \\
& &- \, y_{i}^{th}(z_i,\Omega_M,w(z)) \\
\nonumber \\
S_{i,j} & = &  \delta_{i,j} \sigma_i \sigma_j +
\frac{\partial y_{i}^{th}}{\partial \Omega_M} C_{\Omega_M} 
\frac{\partial y_{j}^{th}}{\partial \Omega_M} + \nonumber \\
& &      \frac{\partial y_{i}^{th}}{\partial w(z)} \cdot \left( C_w 
         \cdot \frac{\partial y_{j}^{th}}{\partial w(z)}  \right) 
\label{s}
\end{eqnarray}

\noindent
To test the accuracy of the inversion we use the {\it a 
posteriori} covariance matrix. It can be shown (see \cite{tarantola2,
nercessian})
that for the linear inverse problem with Gaussian {\it a priori} 
probability density function, the {\it a posteriori} probability density 
function is also Gaussian with mean Eq.\,\ref{solutionMit} and covariance 
Eq.\,\ref{cov}. Although its value is only exact in the linear 
case it is a good approximation here, since the luminosity distance is quite 
linear on the equation of state $w(z)$ at low redshift.

\begin{eqnarray}
\M{C}_{\tilde{\rm{M}}} &=& ( \,\M{G}^*\, \M{C}_y^{-1} \,\M{G} \,+\, 
\M{C}_0^{-1} \,)^{-1} \equiv 
\M{C}_0-\M{C}_0\,\M{G}^*\,\M{S}^{-1}\,\M{G}\,\M{C}_0 \nonumber \\
    &=&  (\,\M{I}-\M{C}_0\,\M{G}^*\,\M{S}^{-1}\,\M{G}\,)\,\M{C}_0
\label{cov}
\end{eqnarray}

\noindent
In an explicit form, the standard deviations from this covariance 
read

\begin{eqnarray}
\label{som}
\tilde{\sigma}_{\Omega_M} &=& \sqrt{C_{\tilde{\Omega}_M}} = \\
&=& \sigma_{\Omega_M}
\sqrt{1- \sum_{i,j}\frac{\partial y_{i}^{th}}{\partial \Omega_M} 
(S^{-1})_{i,j} 
\frac{\partial y_{j}^{th}}{\partial \Omega_M}\sigma_{\Omega_M}^2}\nonumber 
\end{eqnarray}

\begin{eqnarray}
\label{swz}
\tilde{\sigma}_{w(z)}(z) &=& \sqrt{C_{\tilde{w}(z)}(z)} =  \\
&=&\sqrt{ \sigma_{w(z)}^2 - \sum_{i,j} C_{w} \cdot 
\frac{\partial y_{i}^{th}}{\partial w(z)} (S^{-1})_{i,j} 
\frac{\partial y_{j}^{th}}{\partial w(z)} \cdot C_{w}} \nonumber
\end{eqnarray}

\noindent
where the symbols with tilde are the {\it a posteriori} values,
whereas the symbols without represent the {\it a priori} ones.

\begin{figure}
  \begin{center}
    \begin{minipage}[c]{1\linewidth}
        \scalebox{0.58}{\includegraphics{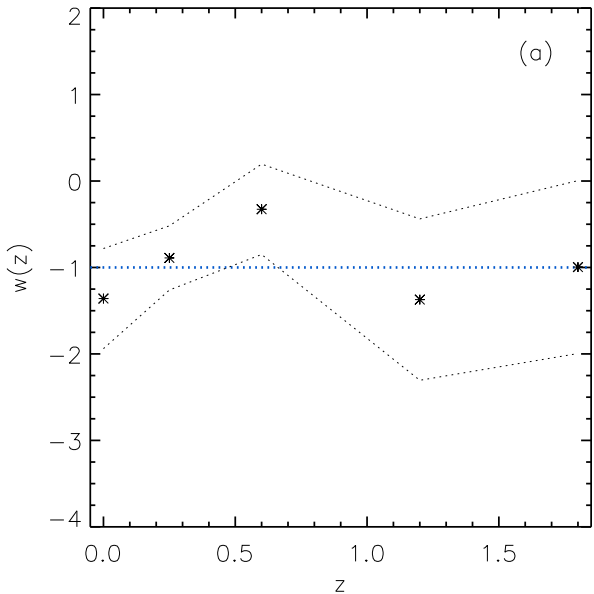}} 
    \end{minipage}
  \end{center}
\caption{Reconstruction of $w(z)$ using  
the 156 SNe from the gold set of Riess et al.(2004) \cite{Riess04}. 
$C_{w(z),w(z')}$ is Gaussian with $\sigma_{w}= 1$ and correlation length
$\Delta_{z} = 0.08$. At high z, the lack of enough data does not allow
to improve the initial knowledge, as seen in
$\tilde{\sigma}_{w(z)}(z) = \sqrt{C_{\tilde{w}(z)}(z)}$.  
Low--resolution inversions (as exemplified  here with $\delta z =$0.3 at low
z, and $\delta z =$ 0.6 at high z) and high
resolution ones (as with $\delta z =$0.06 in Figure 2) give
consistent results.}
\label{fig:wz5points}
\end{figure}

\begin{figure}
  \begin{center}
    \begin{minipage}[c]{1\linewidth}
     \scalebox{0.58}{\includegraphics{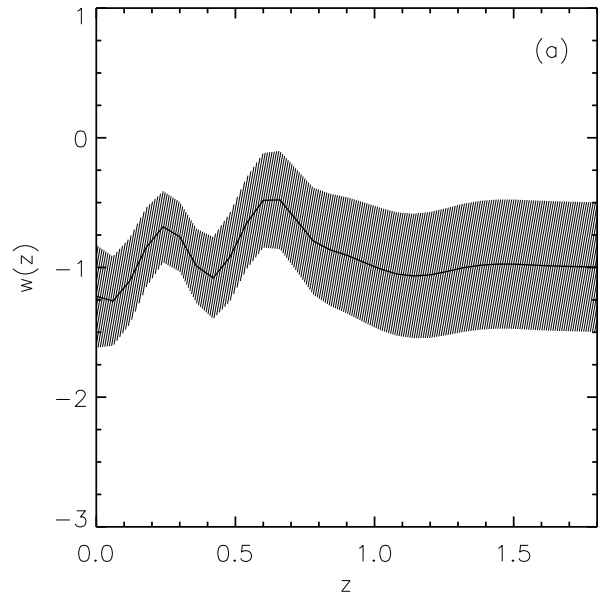}} 
    \end{minipage}\hfill
    \begin{minipage}[c]{1\linewidth}
        \scalebox{0.58}{\includegraphics{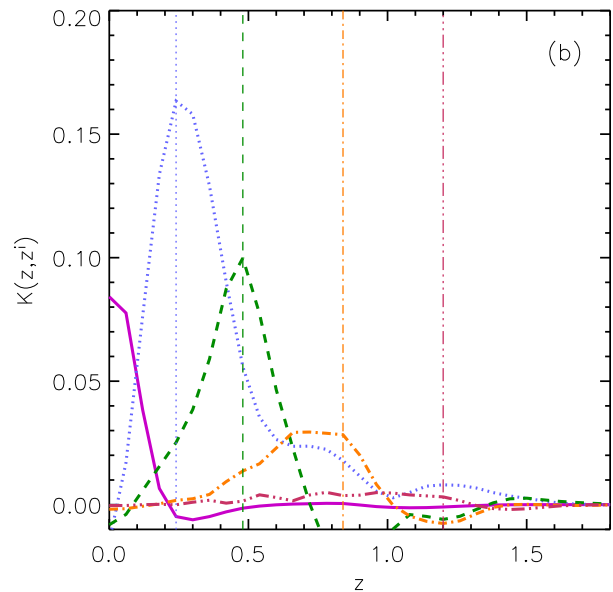}} 
    \end{minipage}
  \end{center}

\caption{Reconstruction of $w(z)$ 
with 156 SNe from the gold set of \cite{Riess04}. These results are obtained 
using Gaussian {\it a priori} covariances with amplitude $\sigma_w=0.5$ and 
$\Delta_z=0.08$. This fine grid calculation uses $\delta$z $=$ 0.06.
The upper panel shows $w(z)$ (solid line) and the 
$1\sigma$ confidence interval (dashed shadow). Below, the different 
resolving kernels at z $=0,0.24,0.48,0.84,1.20$ are shown.
 The resolving kernels at high z show that there is no information to 
conclude on the evolution of the equation of state at z$>$0.6.}
\label{fig:wz156}
\end{figure}

There are other parameters which help to interpret the results.
From the form of Eq.\,\ref{cov} we see that the operator 
$\M{C}_0\M{G}^*\M{S}^{-1}\M{G}$ is related to the obtained resolution. 
This is usually called the 
{\it resolving kernel} $K(z,z^\prime)$. The more this term resembles
the $\delta$-function the smaller the {\it a posteriori} covariance 
function is. In fact, in the linear case, the resolving kernel represents
how much the results of the inversion differ from the true model.
It equals to be the filter between the true model and its estimated
value \cite{backus, tarantola}. In any applied case, it is a
 low band pass filter 
which depends on the data available and the details requested from the model. 
 
 It can also be 
expressed in terms of the {\it a priori} and the {\it a posteriori} 
covariance matrices:
\begin{equation}
 \M{K}\, = \,\M{I}\, - \,\M{C}_{\tilde{\rm{M}}}~\M{C}_0^{-1}
\label{K}
\end{equation}

\noindent 
This expression will be evaluated numerically to quantify the
resolution and information generated in the inversion.
Another term of interest is the  {\it mean index}, which  is
 derived from the resolving kernel:

\begin{equation}
I(z)= \int K(z,z^\prime)dz^\prime.
\label{I}
\end{equation}

\noindent
The nearest $I(z)$  to 
$1$, the most restrictive are the data to the model. 
For very low values, data do 
not improve our prior knowledge on the parameters.

\begin{figure}
  \begin{center}
    \begin{minipage}[c]{1\linewidth}
     \scalebox{0.58}{ \includegraphics{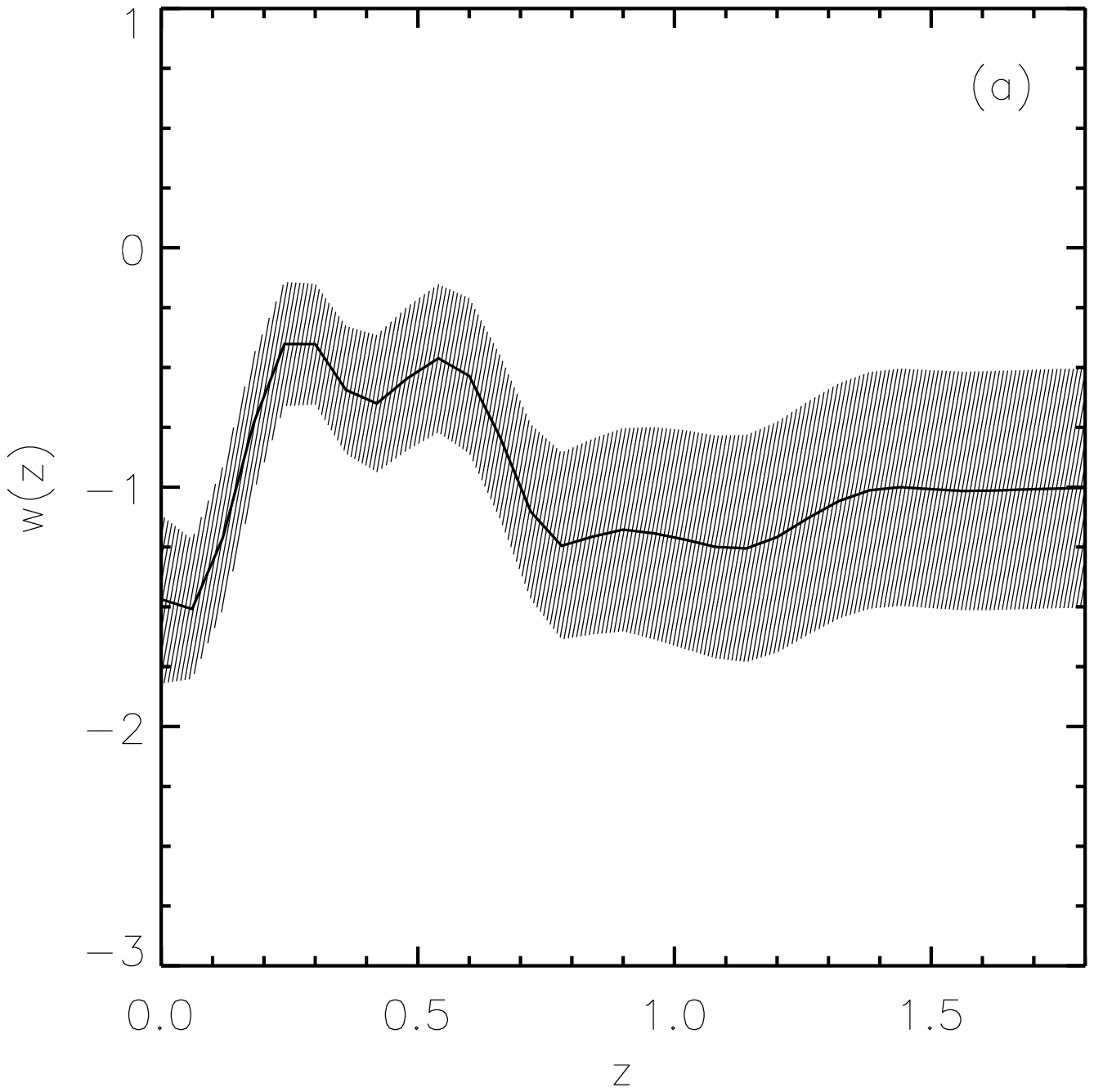}} 
    \end{minipage}\hfill
    \begin{minipage}[c]{1\linewidth}
        \scalebox{0.58}{\includegraphics{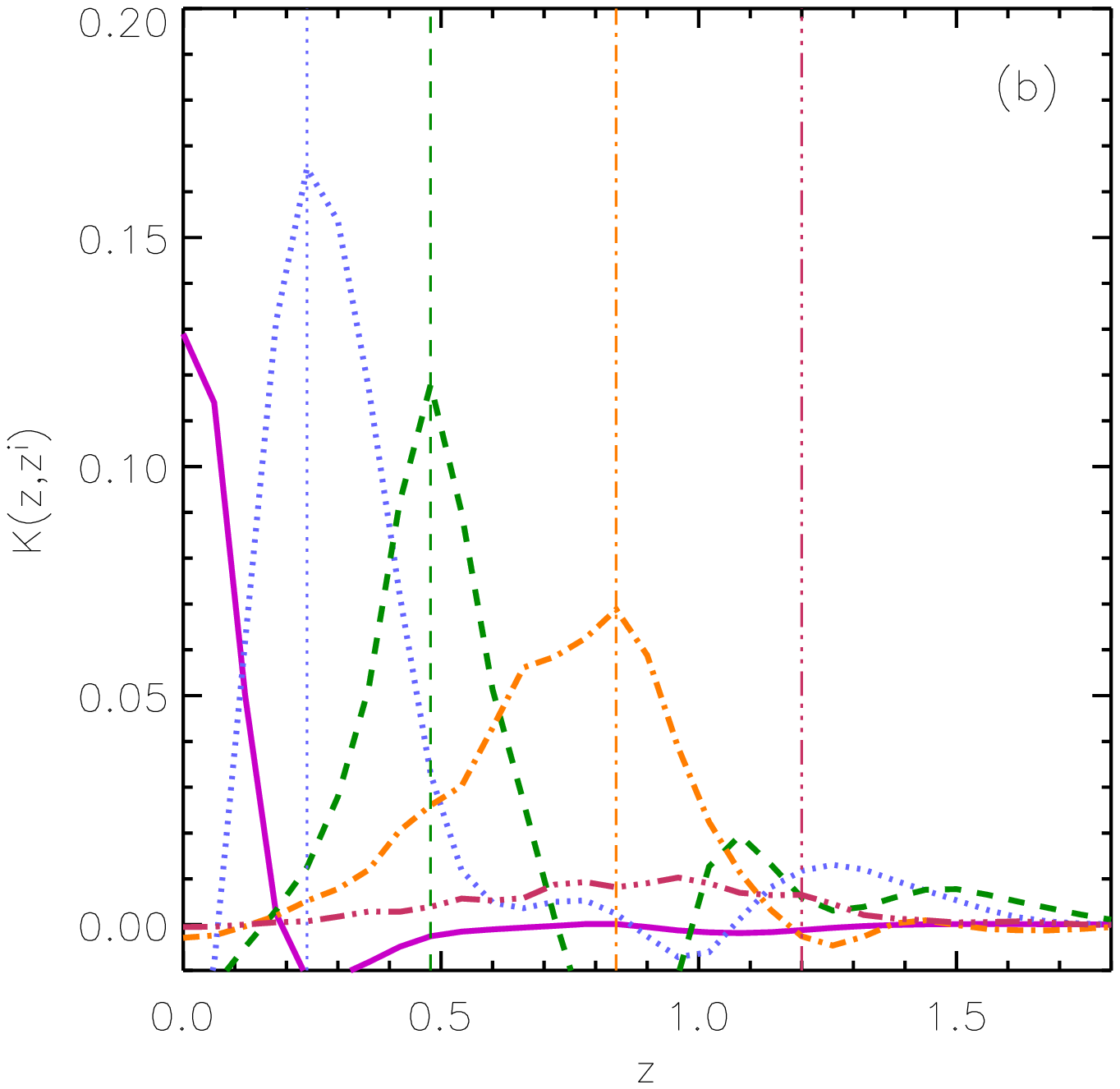}} 
    \end{minipage}
  \end{center}
\caption{Reconstruction of $w(z)$ (as in 
 Figure\,\ref{fig:wz156}) but with the 156 SNe from the 
gold set of \cite{Riess04} and 20 RG from \cite{cdist2}.}
\label{fig:wz176}
\end{figure}

\subsection{Discrete parameters}
\label{discret}

In the previous section we have obtained the results for a set  
of a continuous function and a discrete parameter, 
but we can also consider the case 
of various discrete parameters. It was pointed out that a succesful
parametrization for modeling a large variety of dark energy models
is obtained by considering $w(z)$ expanded around the scale factor
$a$. The earlier parametrization to first order in z given 
by $w(z)= w_0 + w' z$ 
proved unphysical for the CMB data and a poor approach to SN data at z
$\sim$ 1 \cite{Linder2}. 
  For the case of moderate evolution in the equation of state, 
the most simple
(two--parameter) description of $w(z)$ so far proposed is
\cite{polarski, Linder}: 

\begin{equation}
w(z)= w_0 + w_a (1 -a)
\label{EOS0}
\end{equation}

\noindent
where the scale factor $a = (1+z)^{-1}$ and $w(z)$ turns out: 

\begin{equation}
w(z)= w_0 + w_a \frac{z}{1+z}.
\label{EOS1}
\end{equation}

\noindent
 We use now this particular form for the 
function $w(z)$ commonly used to study the behaviour
of dark energy to solve iteratively for  $w_0$ and $w_a$:

\begin{equation}
w_{0[k+1]}= w_{0}^0 + \sigma_{w_0}^2 \sum_{i=1}^N W_{i\,[k]} 
\frac{\partial y_i^{th}}{\partial w_0}_{[k]}
\label{eqw0}
\end{equation}
\begin{equation}
w_{a[k+1]}= w_{a}^0 + \sigma_{w_a}^2 \sum_{i=1}^N W_{i\,[k]} 
{\frac{\partial y_i^{th}}{\partial w_a}}_{[k]}
\label{eqwa}
\end{equation}

\noindent 
where

\begin{equation}
 \frac{\partial y_i^{th}}{\partial w_0}= -\frac{1}{2}
\int_{0}^{z_{i}}\frac{3\Omega_X(z^\prime)\ln(1+z^\prime) dz'}{H^3(z')},
\end{equation}
\begin{equation}
 \frac{\partial y_i^{th}}{\partial w_a}= -\frac{1}{2}
\int_{0}^{z_{i}}\frac{3\Omega_X(z^\prime)[\ln(1+z^\prime)-\frac{z^\prime}
{1+z^\prime}] dz'}{H^3(z')}.
\end{equation}

The {\it a posteriori} variance is for these parameters:

\begin{equation}
\tilde{\sigma}_{w_0} = \sqrt{C_{\tilde{w}_0}} = \sigma_{w_0}
\sqrt{1- \sum_{i,j}\frac{\partial y_{i}^{th}}{\partial w_0} (S^{-1})_{i,j} 
\frac{\partial y_{j}^{th}}{\partial w_0}\sigma_{w_0}^2}
\end{equation}
\begin{equation}
\tilde{\sigma}_{w_a} = \sqrt{C_{\tilde{w}_a}} = \sigma_{w_a}
\sqrt{1- \sum_{i,j}\frac{\partial y_{i}^{th}}{\partial w_a} (S^{-1})_{i,j} 
\frac{\partial y_{j}^{th}}{\partial w_a}\sigma_{w_a}^2}
\end{equation}

\noindent
The equations for $\Omega_M$ are those of section \ref{sec_IP} 
(Eqs.\,\ref{om}, \,\ref{derom} and \,\ref{som}).

\section{Present and Future Samples}
\label{dades}

We briefly introduce the various samples used here for the exploration
of dark energy and for measuring the increase of information in $w(z)$ 
along the last years. 
An initial sample of SNeIa at high z was presented in 1999
by the SCP (Supernova Cosmology 
Project)\cite{p99}. The set includes 16 low-redshift supernovae from 
the Cal\'an/Tololo survey and 38 high-redshift supernovae \cite{p99}. 
The new data of this collaboration
 (low-extinction primary subset in \cite{Knop03}) 
adds 11 high redshift SNe Ia observed with the Hubble Space Telescope 
(HST). The third  SNIa set examined  corresponds to the gold set
buildt up with 7 SNeIa at z $>$ 1 by GOODS and
 a combination of different previous samples 
revised to follow the same calibrations \cite{Riess04}. 
With this last set it has been doubled the maximum redshift and 
triplicated the number of data as compared with the first one. 

\label{wcontinu}
\begin{figure}
  \begin{center}
    \begin{minipage}[c]{1\linewidth}
     \scalebox{0.5}{ \includegraphics{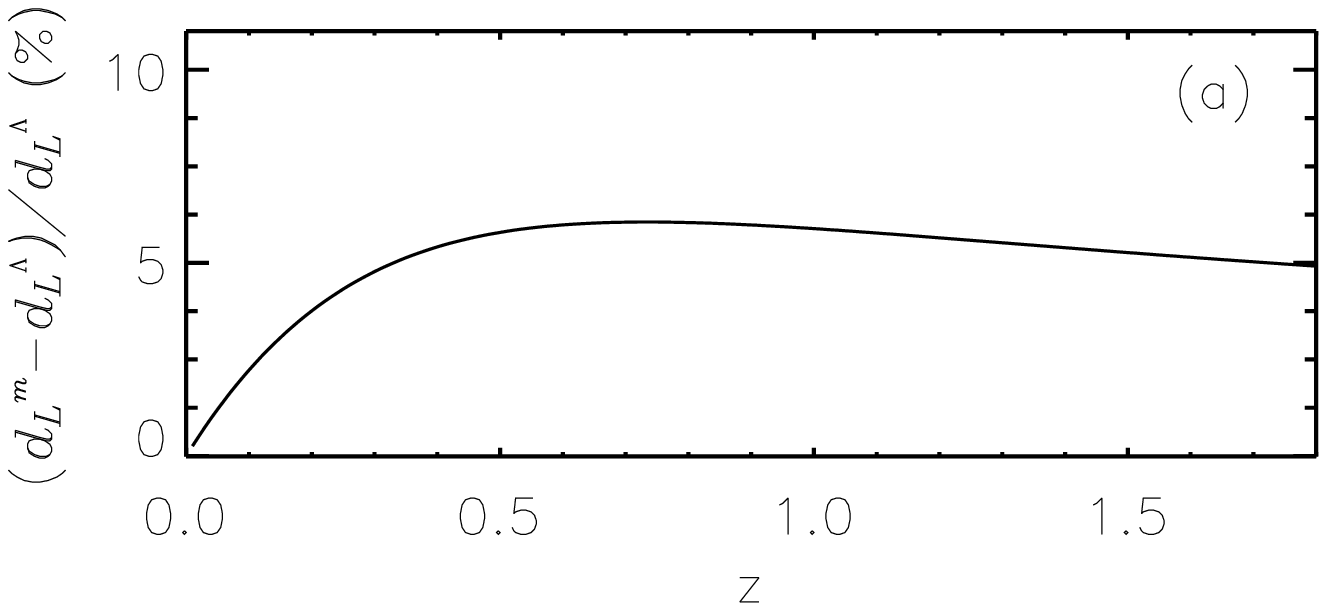}} 
    \end{minipage}\hfill
    \begin{minipage}[c]{1\linewidth}
     \scalebox{0.5}{ \includegraphics{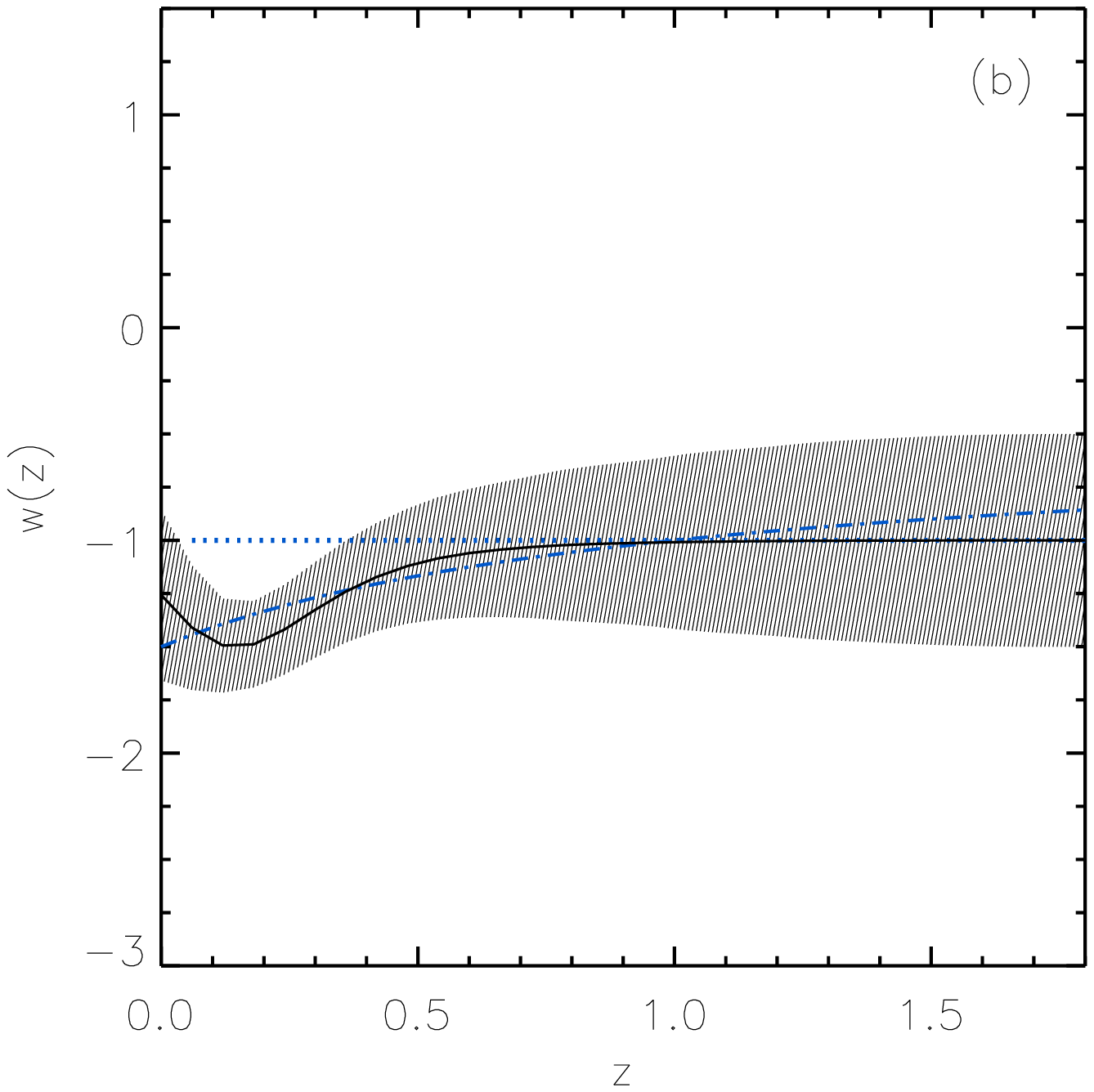}} 
    \end{minipage}\hfill
    \begin{minipage}[c]{1\linewidth}
        \scalebox{0.5}{\includegraphics{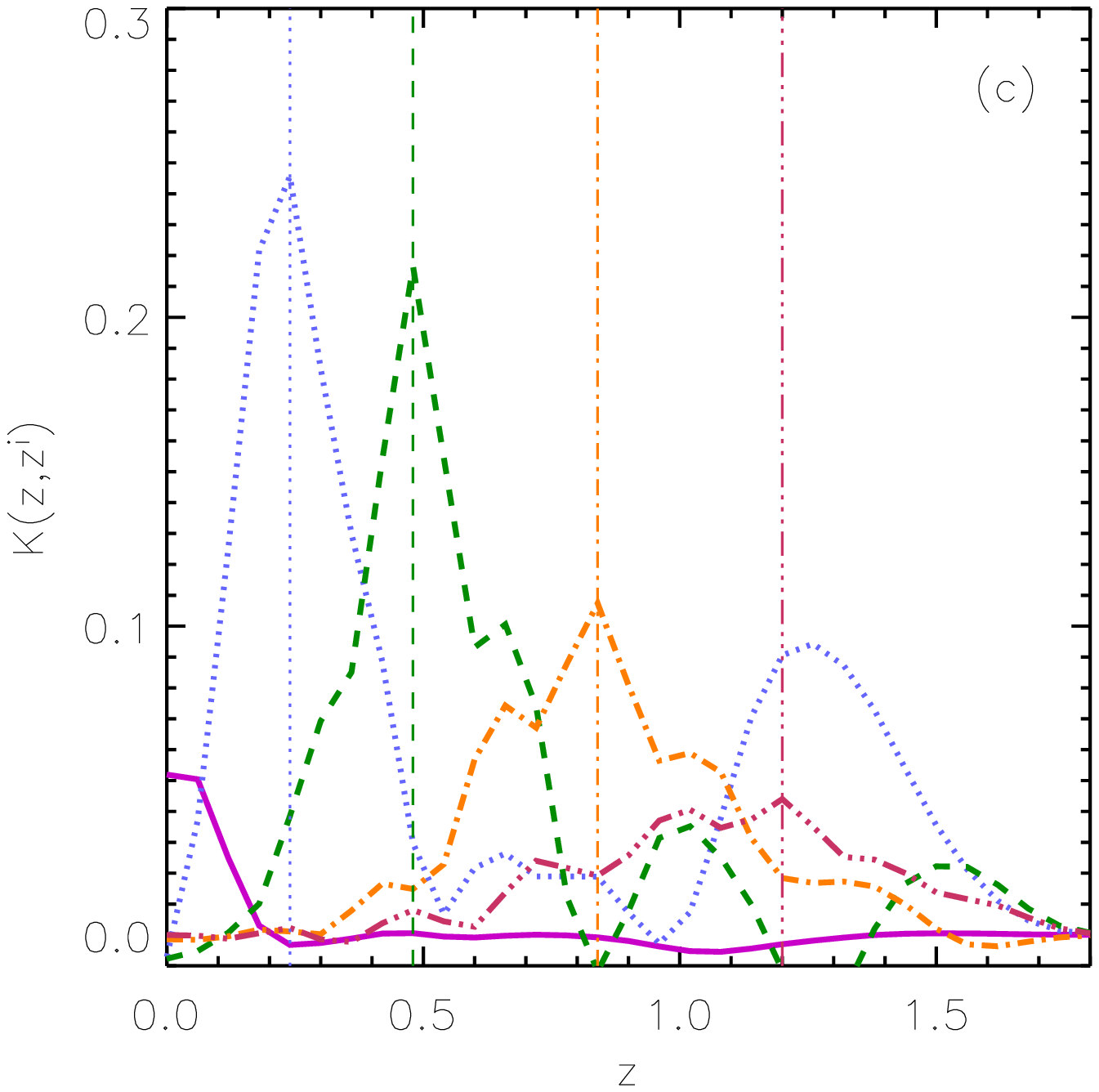}} 
    \end{minipage}
  \end{center}
\caption{Results with 2000 SNe from SNAP plus 300 SNe from SNF. 
 Gaussian covariance is as in Figure \ref{fig:wz156} an 3.
In the upper panel it is shown the deviation in luminosity distance
between model and the cosmological constant
i.e. $(d_L^{m}-d_L^{\Lambda})/d_L^{\Lambda}$. 
The recovered $w(z)$
is shown in the middle panel and the resolving kernels in the low panel. 
The reconstruction,
(shown in solid line in the middle panel) clearly points out that 
at intermediate redshift the best
 model is not the cosmological constant (dotted line), which served as
prior. The inversion recovers the shape of the true dark energy model $w(z)$. 
The dot--dashed line starting at w(0)$=$ --1.5 is the 
fiducial dark energy  model with  $w(z)=-1.5+ 1.0 z/(1+z)$.
The same result is found with a wider prior.  
The fiducial model at z$=$ 0 is below 
$(d_L^{m}-d_L^{\Lambda})/d_L^{\Lambda}$ $=$ 1 $\%$ and away of 
the prior at z=0, but this does not preclude to obtain 
 the slope of the function.}
\label{fig:wzSUGRA}
\end{figure}

\begin{figure}
  \begin{center}
    \begin{minipage}[c]{1\linewidth}
     \scalebox{0.5}{ \includegraphics{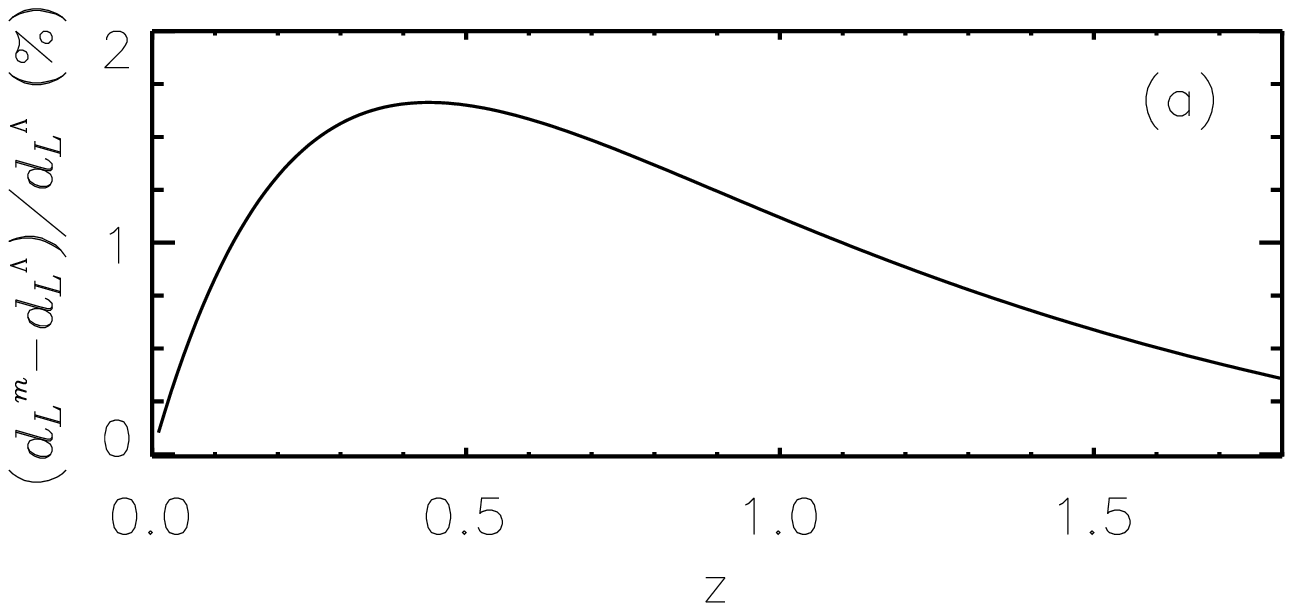}} 
    \end{minipage}\hfill
    \begin{minipage}[c]{1\linewidth}
     \scalebox{0.5}{ \includegraphics{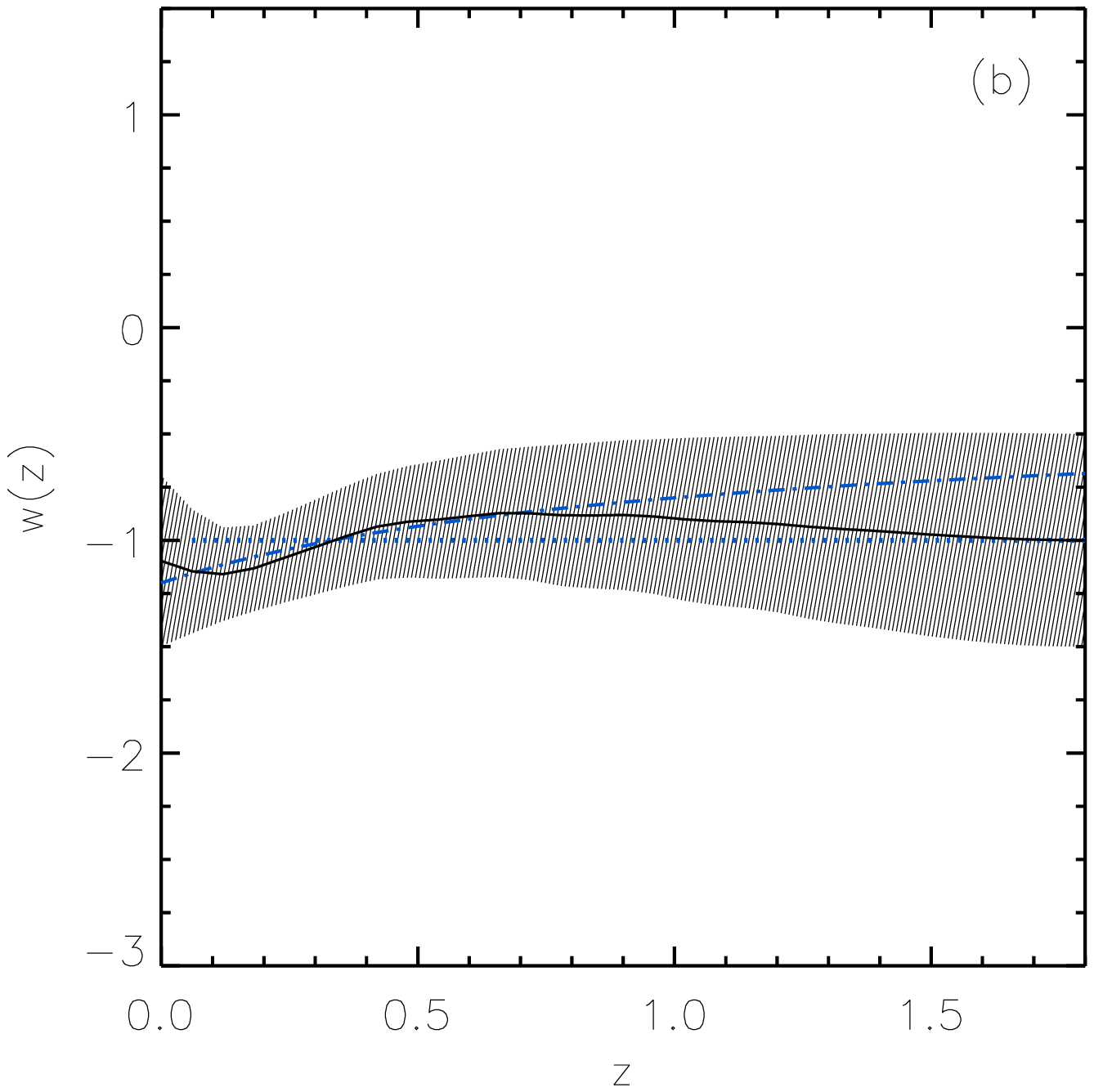}} 
    \end{minipage}\hfill
    \begin{minipage}[c]{1\linewidth}
        \scalebox{0.5}{\includegraphics{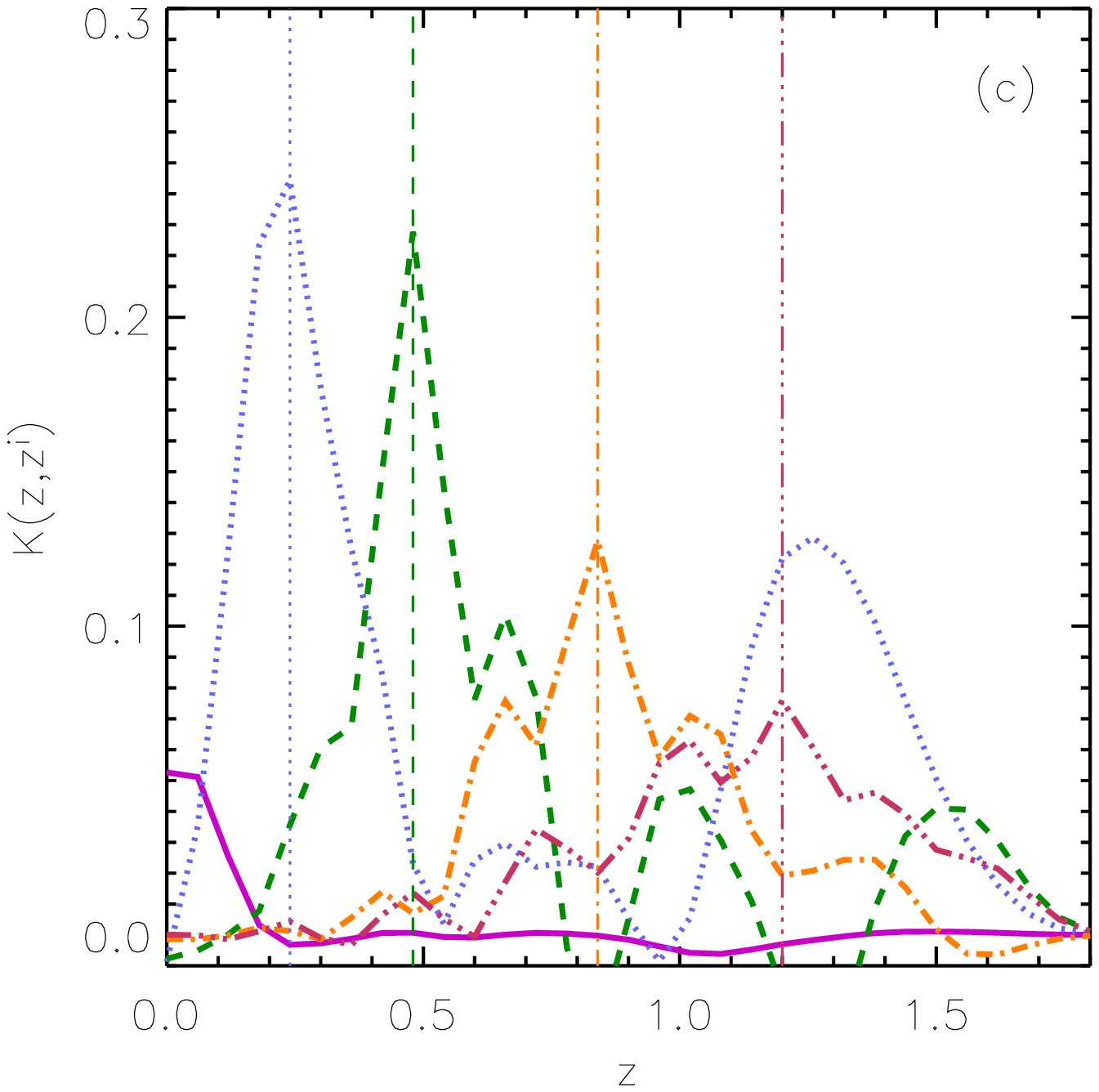}} 
    \end{minipage}
  \end{center}
\caption{Reconstruction of $w(z)$ (as Figure 4), but
now the fiducial model (dot--dashed line)
 is $w(z)=-1.2+0.8z/(1+z)$ and is closer 
in luminosity distance to the prior (dotted line),
 taken to be the cosmological 
constant. In the upper panel it is shown the deviation in luminosity distance
between models, i.e. $(d_L^{m}-d_L^{\Lambda})/d_L^{\Lambda}$.
The true model is recovered at redshifts where the deviation is more
than $1\%$ (a relative difference of 1\% in distance luminosity
corresponds to a magnitude difference between models of $\sim$
0.02 mag,  which at high z is the limiting factor to differentiate
between models). This inversion finds the best solution for models 
which depart
at intermediate z from the cosmological model at more than 1$\%$ in 
$d_L$. This result is independent of the width of the prior.}
\label{fig:wzDEG}
\end{figure}

\begin{figure}
  \begin{center}
    \begin{minipage}[c]{1\linewidth}
        \scalebox{0.58}{\includegraphics{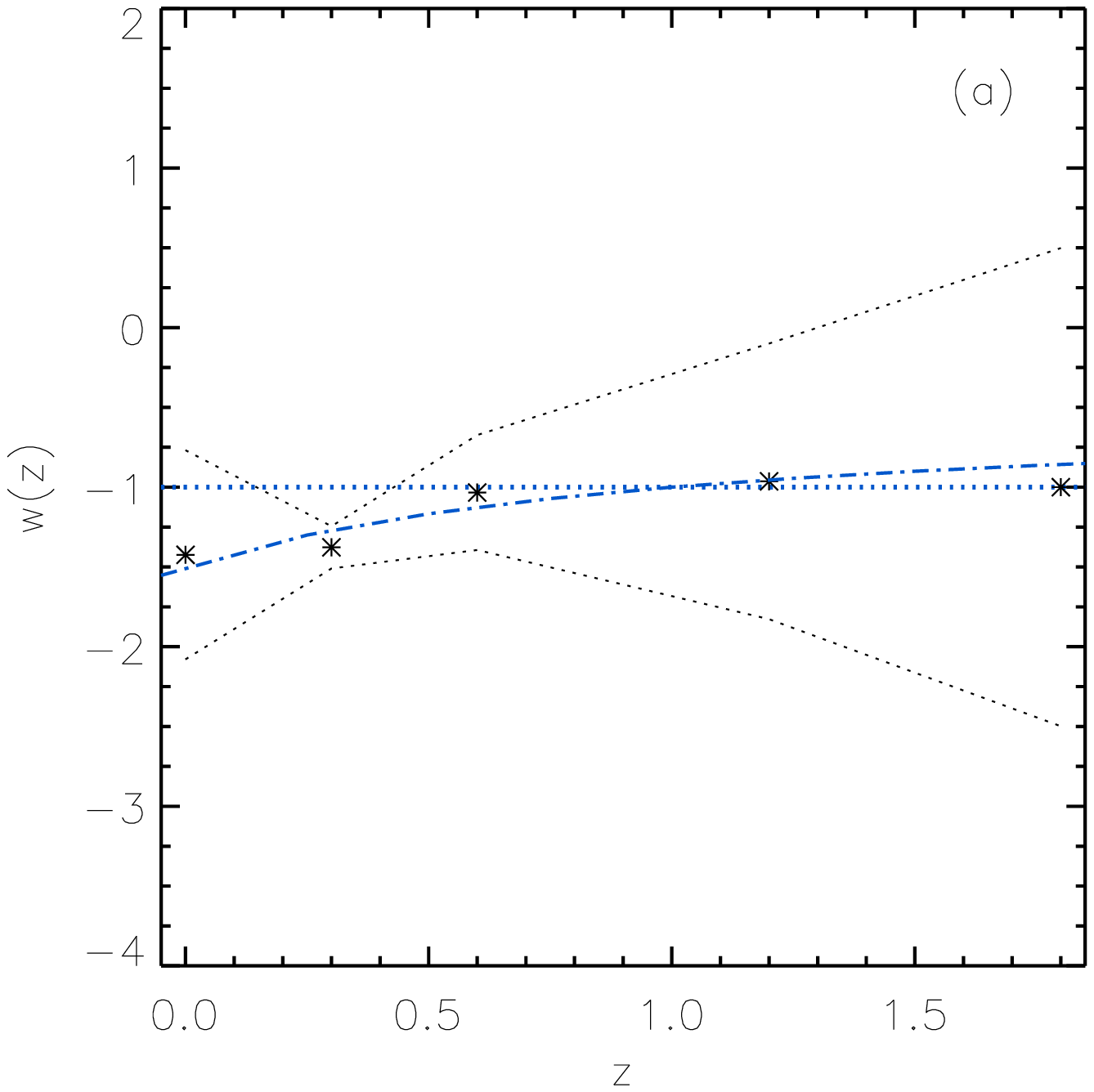}} 
    \end{minipage}\hfill
  \end{center}
\caption{Low resolution reconstruction (5 points calculated as in
Figure 1) of $w(z)$ with 
 the SNAP data set simulated and the fiducial model with 
$w(z)=-1.5+1.0z/(1+z)$. It is shown how this model would be
differentiated from the cosmological constant at more than 2$\sigma$
at intermediate z, in agreement with what is discussed in Figure 4. 
 These results are obtained 
using much wider Gaussian {\it a priori} covariances than in the high
resolution case ($\sigma_w=2.0$). 
}
\label{fig:wz5points}
\end{figure}

The SNIa samples available at present are still small, 
and, therefore,  we also use synthetic samples resembling 
those to be acquired by 
the Supernova Acceleration Probe (SNAP) \cite{snap}. To generate these samples 
 we assume the redshift distribution of 2000 SNeIa from \cite{alexkim}.
For every supernova we 
calculate its magnitude given a fiducial dark energy model. Our first model
corresponds to the cosmological constant in a universe with density parameters 
$\Omega_M=0.3$ and $\Omega_{\Lambda}=0.7$ (this model 
has $w_0=-1$ and $w_a=0$). We generate other synthetic samples based
on dark energy
 models inspired in supergravity theories 
\cite{weller}) with $\Omega_M=0.3$, $w_0$ negative and positive 
$w_a$. Various values of $w_0$ and $w_a$ are tried ($w_0$ range from
--1.5 to --0.8 and $w_a$ range from 0.6 to 1). 
 Gaussian errors are added to the supernova data, 
taking into account  
the statistical error per supernova after the corresponding 
calibrations ($\sigma_{st}=~0.15$). The level of residual systematic
error corresponds to the design of SNAP
\cite{alexkim}, which should achieve 0.02 mag systematic error in
redshifts bins of width $\Delta z$ $=$ 0.1 with a dependence in z,
$\sigma_{sys}= 0.02~z/z_{max}$ \cite{alexkim}. 

\noindent
The error of the measurement in each z bin, is given by: 

\begin{equation}
\sigma_{bin}=\sqrt{\frac{\sum \sigma_{st}^2}{N_{bin}}+\sigma_{sys}^2}
\end{equation} 

\noindent
where $N_{bin}$ is the number of supernovae in each bin. 
To this distribution we also add 300 nearby SNe, as those expected from
the Nearby Supernova Factory \cite{nsf}. A total amount of 2300
SNeIa are used.  
As discussed in \cite{frieman}, with a large data set, 
the irreducible systematic error is
putting the limit to our possibilities of recovering the dark energy
equation of state.

At the moment, the
low numbers of SNe Ia found at z$>$ 1, 
impact in the reconstruction of the
equation of state.  
Although SNe\,Ia are the best known calibrated candles at high 
redshift, we can explore other luminous sources
that could be reliable distance indicators
to study dark energy.  Among these, the 
Fanaroff--Riley type IIb radio galaxies (FR IIb) are a 
group of galaxies quite homogeneous 
along redshift \cite{dalyRG}. Their angular size measured
from the outer edges of their lobes can be used to determine angular
distances up to high z. Their possible evolution along z  
 and selection effects have been
studied \cite{dalyRG}. We use the 20 dimensionless 
coordinate distances to the 
FR IIb radio galaxies available \cite{cdist2}. 
This sample extends to z $=1.8$, thus
 complements the gold set of SNe\,Ia 
\cite{Riess04} at high redshift.

We also consider a third type of sources, compact radio sources (CRS).
Extended objects of this kind have not yet been fully tested for  
evolutionary effects. However,  a sample of 330 sources is available
and they provide distances to high z
\cite{gurvits}. Here we use the subset with spectral index $-0.38\leq
\alpha\leq0.18$ and total radio luminosity $Lh^2\geq10^{26}~W/Hz$ to minimize
the possible dependences in angular size-spectral index and linear
size-luminosity \cite{gurvits}. As done in this last reference and some
other subsequent analysis \cite{lima,zhu}, the 145 data have been binned in 12 
intervals with 12 or 13 sources each, from z $=0.52$ to z $=3.6$. 
In order to fit exclusively the parameters appearing in the equation of state
 a value of the characteristic length must be adopted. We use that obtained 
in the best fit of \cite{lima} and \cite{zhu}, $l=22.64h^{-1}pc$. 

Other kinds of objects such as core--collapse SNe \cite{collapse2} or 
gamma ray bursts \cite{grbparam} have been proposed as cosmic distance
indicators observable up to very high z.
 Large samples of those
distance indicators with well studied errors 
are lacking, and for those reasons they have not
 been explored in this study of dark energy.

\section{Determination of $\lowercase{w(z)}$}

We determine  $w(z)$ using the
inverse approach described above.  
Explicitly, one obtains the value of $\Omega_M$ and $w$ 
at a given redshift with the equations of section \ref{sec_IP} 
(Eqs.\,\ref{om}, \ref{wz}, \ref{som} and \ref{swz}).

The {\it a priori} model is arbitrary, but determines where the solution
is searched. In the situation where the data are scarce  with very 
wide priors in the function
$w(z)$, the solution can iterate between saddle points and local
minima as few data do provide a landscape with no strong minima.
Fortunately, in our case, the sample available for $w(z)$ 
is large enough to allow to find the solution with good reliability. 
The solution  using the 156 gold set \cite{Riess04} is shown in Figure 1.
We tried different
priors and looked at the regions where $w(z)$ could be found. 
We found a solution in the area of negative $w(z)$ 
 and no solution outside an area of 
$\sigma_{w}$ = 1 around  $w(0)$$=$ --1.  Being a Gaussian prior with
$\sigma_{w}$ $=$ 1, if there would have been a solution peaking
strongly towards positive $w(z)$ at intermediate z, it would have
been found, as shown in our exploration. We are very unrestrictive in 
how fast $w(z)$ can go with z as we keep a correlation 
length allowing for fast changes of slope.

Other priors were tried to check for the stability of 
the solution found.  Our results are stable against the width of
prior. The solution with 
$\sigma_{w}$ $=$ 0.5 is found in Figure 2.  
The black solid line indicates the evolution of 
the equation of state, whereas the shadowed region represents the 1$\sigma$ 
interval.  The low resolution results show the 
same trend. 
Choosing a more peaked prior model (stronger prior), allows to explore 
solutions in a narrower range and recover them with more resolution. 
The solutions found are similar independently of the
choice of covariance function. 
Various functional forms for the covariance
have been tried giving the same results (covariance as 
in Eq.\,\ref{covwexp} or Gaussian give similar results).  
The number of 
iterations needed to converge, i.e,   $|w_k - w_{k-1}|< \epsilon$, with 
$\epsilon$ of a few percent, is typically less than 10.


The results of Figure 1 and 
Figure \ref{fig:wz156}
 are compatible with a cosmological constant already 
 at about 1$\sigma$ level. The ascent of $w(z)$ towards $0$ 
at intermediate redshift appears with SNe Ia and radiogalaxies, though
not at a high significance level.   
The two peaks in Figure 2 are only seen in high resolution and are not 
significant as shown by $\tilde\sigma_{w(z)}$. 
The resolving kernels 
(low panel of the Figure \ref{fig:wz156}) indicate that the function
is generally not well resolved at individual redshifts, well beyond 
the redshift range z $\sim 0.5$. At high 
redshift, z $=1.2$ for example, we observe a very wide and extremely flat 
$K(z,1.2)$, meaning that this redshift is not resolved at all by the
data. The reliability of the inversion peaks in the range of z $\sim$ 
 0.2--0.5 where the information is maximum.

Using the combination of  
SNe\,Ia and FR IIb radiogalaxies implies to add 
20 additional objects, but almost all of them 
are located at high redshift. This translates 
into a slightly better resolving kernels up to z $\sim 0.6$ as
 it can be seen in 
Figure \ref{fig:wz176}. The equation of state still shows 
the peak approaching to zero at intermediate redshifts. 
However, from the data gathered up to now
 there is no information to infer that the positive trend from 
$w_{0}$ $<$ --1 to 
an increase up to 0 at z $\sim$  0.2--0.5 tentatively seen 
continues beyond z $>$ 0.6. In our approach, at high
redshift, where there are no data, the method recovers the prior. With 
this larger set of data, the cosmological constant is  at all z
within the 2$\sigma$ contour. We find similar results as in
\cite{cdist2,Huterer3}.

In our iterative procedure, $\chi^2$ is estimated at each iteration
as \cite{tarantola}: 

\begin{equation}
\chi^2_{[k]} = (\M{y}-\M{y}^{th}(\M{M}_{[k]}))^* ~ \M{C}_y^{-1} ~
(\M{y}-\M{y}^{th}(\M{M}_{[k]}))
\label{chi2k}
\end{equation}

A simple description of the goodness of the fit is given by $\chi^2_{\nu} = 
{\chi^2} / N_{data}$, being close to 1 in a good fit. We have at
convergence for the SNeIa sample (Figures 1 and 2) $\chi^2_{2}$=123,
and $\chi^2_{\nu}$ $=$ 0.72, while for the combined sample of 
of SNeIa and radiogalaxies  $\chi^2$=214 and $\chi^2_{\nu}$ $=$
1.21.

The power of the method and the facility to find the best estimate for
$w(z)$ largely improves for samples such as those to be
gathered by SNAP, where the prior can be set to be widely 
uninformative. 
We tested the 
method's capability to recover the equation of state using the synthetic data 
of SNAP simulated samples. 
The method can reconstruct solutions 
which are degenerate with the cosmological constant at some z, having  a
 relative difference in luminosity distance
  $\Delta d_L(z)$ lower than 1$\%$, but differing from it
at other z. 
In Figure 4, we have plotted (short 
dashed line) the equation of state of the fiducial dark energy model 
$w(z)=-1.5+1.0z/(1+z)$. 
The reconstruction has been overplotted together with 1$\sigma$ uncertainties, 
and it can be seen that we obtain an improvement of the prior
at intermediate redshift. The reconstruction differs from the
cosmological constant at more than 2$\sigma$ (see also Fig 6 for
an additional discussion). 
The resolving kernels also show that the intermediate z is the best
resolved redshift range. 
As it happens in all data sets, 
the redshift z $=0$ is worse determined than higher
z. Multiple peaks appear in the resolving kernel and the kernel is
wide at some z, which results in a degree of degeneracy of the
function at those
redshifts. This is a result to be expected as the dependence between 
the equation of state and the luminosity distance ($w(z)$ is hidden within a
double integral in redshift, and thus, 
its variation with redshift is smoothed) eludes  
the uniqueness of the result. 
 
 In Figure \ref{fig:wzDEG}, our synthetic sample 
corresponds to an equation of state $w(z)=-1.2 + 0.8 z/(z+1)$. 
 The corresponding 
difference in the luminosity distance between both models is less than a 
$2\%$, and despite this,  the method is able to recover the real model at 
intermediate redshift, where the degree of information in $w(z)$
 is higher but also where
this luminosity distance difference is larger than $1\%$.
 At high
redshift, where there are fewer data and very small deviations in the
luminosity distance, the prior is not improved. 
There, the SNAP
sample meets its systematic error of 0.02 mags, and the
limit of discernibility of the models is shown in the impossibility to
find variations in $w(z)$ implying  less than 
1$\%$ in $\Delta$d$_{L}$. 
 However,  we can discriminate  models against the cosmological
constant which are 1$\%$ above de discernibilty at some z, while
they might fall below it at other z. 
This result is found at all resolutions. 
Figure 6
is exactly the same case as Figure 4, but 
with the wider prior ($\sigma_w=2.0$).

\section{Information on $\lowercase{w}_0$ and $\lowercase{w_a}$}
\label{w0wa}

The results for $\Omega_M$, $w_0$ and 
$w_a$ for different {\it a priori} models and for the SNe samples  
 can be compared with what is
obtained in the non--parametric reconstruction of $w(z)$. A summary 
for the results using SNeIa samples is given in  Table\,\ref{taula}. 
The three discrete unknown parameters are $\Omega_M$, $w_0$ 
and $w_a$.
We apply
uninformative priors, i.e. large covariances. 


\begin{table}[h]
\caption{Priors and results for $\Omega_M$, $w_0$ and $w_a$. Values in 
parentheses reflect the error of the parameter in the last digit. To 
see the reliability of the result the mean index,
$I$, is shown. Imposing a prior --1(0) for $w_0$ means that we force the result
to be cosmological constant now. Alternatively, a prior of 
 $w_a$ with value 0(0) forces no evolution. Certainly, the most
interesting case is when priors are wide: $\Omega_M$ $=$ 0.30(4),
 $w_{0}$ $=$ --1(10) and $w_{a}$ $=$ 0(10) and there are no
restrictions (results outlined in boldface).
}
\label{taula}
\begin{center}
\begin{ruledtabular}
\begin{tabular}{c c c c c c c} 
     &           & &           & &          &                    \\
     &$\Omega_M$&$I_{\Omega_M}$&$w_0$&$I_{w_0}$&$w_a$&$I_{w_a}$  \\
     &           & &           & &          &                    \\
\hline
     &           & &           & &          &                    \\
{\it Prior}& ${\it 0.30(0)}$ & {\it -}  & ${\it -1(10)}$ & {\it -} & ${\it 0(0)}$ &{
\it -}   \\
     &           & &           & &          &                    \\
P99 & $0.30(0)$ &-&$-1.0(2)$ &0.999& $0(0)$ &-\\
K03 & $0.30(0)$ &-&$-1.1(1)$ &0.999& $0(0)$ &-\\
R04 & $0.30(0)$ &-&$-1.0(1)$ &0.999& $0(0)$ &-\\
SNAPcc& $0.30(0)$ &-&$-1.00(4)$ &0.999& $0(0)$ &-\\
SNAPsu& $0.30(0)$ &-&$-0.69(3)$ &0.999& $0(0)$ &-\\
     &           & &           & &          &                    \\
\hline
     &           & &           & &          &                    \\
{\it Prior}& ${\it 0.30(4)}$ & {\it -}  & ${\it -1(10)}$ & {\it -} & ${\it 0(0)}$ &{
\it -}   \\
     &           & &           & &          &                    \\
P99 & $0.30(3)$ &0.245&$-1.0(2)$ &0.999& $0(0)$ &-\\
K03 & $0.31(3)$ &0.316&$-1.1(2)$ &0.999& $0(0)$ &-\\
R04 & $0.29(3)$ &0.540&$-1.0(2)$ &0.999& $0(0)$ &-\\
SNAPcc& $0.30(3)$ &0.344&$-1.0(1)$ &0.999& $0(0)$ &-\\
SNAPsu& $0.28(3)$ &0.432&$-0.7(1)$ &0.999& $0(0)$ &-\\
     &           & &           & &          &                 \\
\hline
     &           & &           & &          &                 \\
{\it Prior}& ${\it 0.30(4)}$ & {\it -}  & ${\it -1(10)}$ & {\it -} & ${\it 0(10)}$ &{\it -}
\\
     &           & &           & &          &                 \\
{\bf P99} & {\bf 0.30(4)} & {\bf 0.193} & {\bf -1.5(7)} & {\bf 0.995} & {\bf +4(5)} & {\bf 
0.739}\\
{\bf K03} & {\bf 0.31(3)} & {\bf 0.254} & {\bf -1.1(7)} & {\bf 0.995} & {\bf +0(5)} & {\bf 
0.774}\\
{\bf R04} & {\bf 0.31(3)} & {\bf 0.495} & {\bf -1.3(4)} & {\bf 0.998} & {\bf +2(2)} & {\bf 
0.951}\\
{\bf SNAPcc} & {\bf 0.30(3)} & {\bf 0.334} & {\bf -1.0(3)} & {\bf
0.999} &
 {\bf +0(1)} & {\bf 0.987}\\
{\bf SNAPsu} & {\bf 0.30(3)} & {\bf 0.314} & {\bf -0.8(2)} & {\bf 0.999} & {\bf +0.6(9)} & 
{\bf 0.991}\\
    &           & &           & &          &                 \\
\hline

     &           & &           & &          &                 \\
{\it Prior}& ${\it 0.30(4)}$ & {\it -}  & ${\it -1(0)}$ & {\it -} & ${\it 0(10)}$ &{\it -}\\
     &           & &           & &          &                 \\
P99 & $0.29(3)$ &0.383&$-1(0)$ & - & $+0(2)$ &0.975\\
K03 & $0.30(3)$ &0.460&$-1(0)$ & - & $-1(2)$ &0.974\\
R04 & $0.27(2)$ &0.669&$-1(0)$ & - & $+0.6(8)$ &0.994\\
SNAPcc& $0.30(2)$ &0.645&$-1(0)$ & - & $+0.0(7)$ &0.999\\
SNAPsu& $0.35(2)$ &0.721&$-1(0)$ & - & $+1.2(5)$ &0.997\\
     &           & &           & &          &                 \\

\end{tabular} 
\end{ruledtabular}
\end{center}
\end{table}

The first sample, quoted as {\it P99}, 
corresponds to the data used in the main fit of \cite{p99}. The second one, 
{\it K03} represents the low-extinction primary subset of \cite{Knop03}, 
whereas {\it R04} is the gold set of \cite{Riess04}. {\it SNAPcc} is the 
SNAP simulation for the CC model and {\it SNAPsu} corresponds to the SUGRA
model with $w_{0}$ $= -0.8$, $w_{a}$ $= +0.6$. 
 We use Eqs.\,\ref{om}, \,\ref{eqw0} and \,\ref{eqwa} and 
make all the numerical calculations
as in the previous section.

The mean index
obtained in the determination of $w_{0}$ and $w_{a}$ gives us an
indication of the information contained in each sample of data, and
how much improvement is obtained when adding more SNe Ia or other 
distance indicators covering wider z ranges. 
 It shows whether the result is reliable 
($I\sim1$), and whether is highly dependent on the {\it a priori} model 
($I<<1$). 

Several situations are examined to
size the effects of  our prior knowledge in the determination of
$w_a$. If we would know independently $w_0$, the derivative $w_a$, would be
determined with high reliability. 
 It is possible that
this could be done with some other method providing a prior on $w_0$.   
In the case of a complete ignorance of $w_0$, a good degree of knowledge 
of $\Omega_M$ also helps to determine $w_a$. The error decreases by a
factor 2 when we go from $\sigma_{\Omega_M}=0.04$ to
$\sigma_{\Omega_M}=0$.

The mean 
index obtained for $w_{0}$ and $w_{a}$ is relatively high and
increasing with the large SNAP sample. The low mean index found  for 
$\Omega_M$ just indicates
that we had from the beginning a good knowledge on the parameter
and the data can not improve a lot its uncertainty although the fit is
good.

The linear expansion in $w(z)=w_0+w'z$ is used in 
Riess et al. (2004) \cite{Riess04}. In our inversion, we found that 
a linear expansion in $w(z)=w_0+w'z$ produces low reliability indexes
as it incorporates very poorly the high redshift information.
High reliability indexes are found in the expansion in terms of 
$w_0$ and $w_a$. 
With the inverse method, we find: $w_0=-1.3\pm0.4$ and $w_a= 2 \pm 2$.
 When using FRIIb and SNe Ia, we have a similar
 result with half the error  $w_a= 2 \pm 1$, in consistency with what
is found in the non--parametric analysis.

\begin{table}[h]
\caption{The same as Table\,\ref{taula} but now priors and results are for 
 models of evolving CC with parameters $\Omega_M (=1-\Omega_\Lambda)$ and 
$(1/\rho_c^0)d\Lambda/dz|_0$.}
\label{taulalambda}
\begin{center}
\begin{tabular}{c c c c c} 
\hline
\hline
       &           & &           &                               \\
       &~~~~~~$\Omega_M$~~~~~~&~~~$I_{\Omega_M}$~~~&~~~$\left.\frac{1}{\rho_c}\frac{
d\Lambda}{dz}\right|_0$~~~&~~~$I_{d\Lambda}$~~~     \\
       &           & &           &              \\
\hline

     &           & &           &              \\
{\it Prior}& ${\it 0.30(4)}$ &{\it -}  & ${\it 0(10)}$ &{\it -}  \\
     &           & &           &              \\
{\bf P99} & {\bf 0.30(3)} &{\bf 0.262} & {\bf 0.0(4)} &{\bf 0.998}\\
{\bf K03} & {\bf 0.31(3)} & {\bf 0.352} & {\bf -0.2(3)} &{\bf 0.998}\\
{\bf R04} & {\bf 0.29(3)} & {\bf 0.555} & {\bf 0.0(3)} &{\bf 0.999}\\
{\bf SNAPcc}& {\bf 0.30(3)} &{\bf 0.389} & {\bf-0.0(2)} &{\bf 0.999}\\
{\bf SNAPsu}& {\bf 0.29(3)} &{\bf 0.371} & {\bf+0.7(3)} &{\bf 0.999}\\
     &           & &           &              \\
\hline
     &           & &           &              \\
{\it Prior}& ${\it 0.30(10.0)}$ & {\bf- } & ${\it 0(10)}$ & {\it -} \\
     &           & &           &              \\
P99 & $0.28(7)$ &0.999&$+0.1(7)$ &0.995\\
K03 & $0.32(5)$ &0.999&$-0.4(5)$ &0.997\\
R04 & $0.29(4)$ &0.999&$+0.1(4)$ &0.998\\
SNAPcc& $0.30(5)$ &0.999&$-0.0(4)$ &0.998\\
SNAPsu& $0.28(5)$ &0.999&$+0.8(4)$ &0.998\\
     &           & &           &              \\

\hline 
\hline
\end{tabular} 
\end{center}
\end{table}

Some models of dark energy are not well parameterized
using an equation of state with $w_0$ and $w_a$.
Here we investigate how this inverse approach works for the 
varying cosmological constant models. The background of those models
is that quantum effects near 
the Planck scale would cause the evolution of the CC
 (see \cite{babic,article}). The best way to parameterize the dark 
energy density along z representing
 all this branch of models is through:
 
\begin{equation}
\Omega_X(z)= \Omega_\Lambda^0 + 
\left.\frac{1}{\rho_c^0}\frac{d\Lambda}{dz}\right|_0z
\label{derlam}
\end{equation}

Thus, we have implemented the inverse approach for this branch of
 models and find the results for the two discrete parameters that are
to first order describing this dark energy candidate. Those are 
$\Omega_M(=1-\Omega_\Lambda)$ and $(1/\rho_c^0)d\Lambda/dz|_0$. Results are 
shown in Table\,\ref{taulalambda}. Although present-day data are 
consistent with a non-variation of the CC, a small running is allowed.
These results can be compared with those from \cite{article}, where
similar conclusions were reached through a $\chi^2$-test analysis.

The simulation using a SUGRA fiducial model with $w_{0} =-0.8$ and
$w_{a} =+0.6$ as observed by SNAP, shows that the data set
 is degenerate with a $\Lambda$--varying model with 
$(1/\rho_c^0)d\Lambda/dz|_0$ $= +0.7$. A $\Lambda$--varying model such
as those proposed \cite{babic} results 
in observables that, if interpreted through $w_{0}$ and $w_{a}$,
might suggest a model with quite different physics.

Overall, degeneracies found in the solution of $w(z)$ and the
existence of models 
apparently indicating an evolving $w(z)$ but reflecting other
physics,  will make necessary to analyse the set of possible
 theories compatible with SNAP observations. 


\begin{table}[h]
\caption{Priors, results and mean index for $\Omega_M$, $w_0$ and $w_a$ but
for different kinds of sources: SNeIa gold set (SN), FR IIb 
radio galaxies (RG), compact radio sources (CRS) and combinations among them.}
\label{taulasources}
\begin{center}
\begin{ruledtabular}
\begin{tabular}{c c c c c c c} 
     &           & &           & &          &                    \\
     &$\Omega_M$&$I_{\Omega_M}$&$w_0$&$I_{w_0}$&$w_a$&$I_{w_a}$  \\
     &           & &           & &          &                    \\
\hline

     &           & &           & &          &                 \\
{\it Prior}& ${\it 0.30(4)}$ & {\it -}  & ${\it -1(10)}$ & {\it -} & ${\it 0(10)}$ &{\it -}\\
     &           & &           & &          &                 \\
{\bf SN} & {\bf 0.31(3)} & {\bf 0.495} & {\bf -1.3(4)} & {\bf 0.998} & {\bf +2(2)} & {\bf 0.951}\\
{\bf RG} & {\bf 0.30(4)} & {\bf 0.193} & {\bf 0(1)} & {\bf 0.988} & {\bf -6(7)} & {\bf 0.492}\\
{\bf CRS} & {\bf 0.30(4)} & {\bf 0.015} & {\bf 0(1)} & {\bf 0.984} & {\bf -3(7)} & {\bf 0.518}\\
{\bf SN+RG} & {\bf 0.28(2)} & {\bf 0.640} & {\bf -1.3(3)} & {\bf 0.999} & {\bf +2(1)} & {\bf 0.979}\\
{\bf SN+CRS} & {\bf 0.28(2)} & {\bf 0.642} & {\bf -1.4(3)} & {\bf 0.999} & {\bf +2(1)} & {\bf 0.982}\\
{\bf all} & {\bf 0.27(2)} & {\bf 0.647} & {\bf -1.3(3)} & {\bf 0.999} & {\bf +2(1)} & {\bf 0.984}\\
     &           & &           & &          &                   \\

\end{tabular} 
\end{ruledtabular}
\end{center}
\end{table}

Combining SNe Ia  with the other sources introduced in 
section \ref{dades} will increase considerably the number of data at very high
redshift.

As can be seen in Table \ref{taulasources}, 
using the three kinds of cosmic distance indicators reduces the
uncertainty in the first derivative of the equation of state $w_{a}$ 
 by 50$\%$ in respect to the use of SNeIa.
Although data from CRS extend 
to z $=3.6$, we obtain similar results by only adding 20 FRIIb
radiogalaxies which reach up to
z $=1.8$. 

As a difference with the gold set of \cite{Riess04} (results in Table
 \ref{taula}) we observe a positive evolution of the equation of state 
($w_a>0$) at almost 2$\sigma$ level when more than a set of data is 
considered. This might bring again
the need to look for constraints in $w(z)$ at very high z, and to
examine closely the systematic effects of cosmic distance indicators. 
Different samples of
distance indicators might favor slightly different results.
As seen in Fig. 2 and Fig. 3, evolution 
is only tentatively suggested at z $\sim$ 0.3--0.5. 
Moreover, trends of evolution at high z can not be determined with the
 available samples. Those  redshifts are very poorly
determined.

\section{Summary and discussion}
\label{summary}

We introduce  here an 
Inverse Problem approach to determine $w(z)$ as a continuos 
function in a model--independent and non--parametric way. 
The method retrieves $w(z)$ without imposing any constraints in 
the form of the function. 
The method uses Bayesian information such as 
 the area where this solution is to be found, which can be quite
unrestrictive, as shown with simulations using synthetic SNe Ia samples of the 
size and systematic errors of SNAP.
In a situation with low amount of data, the algorithm can become
unstable if the solution is to be found through a very large area. 
Constraining then the a priori information,  helps to stabilize the
solution, but a careful exploration for the presence of other
solutions changing the prior should be done. 
In fact, lowering the covariance of the prior in $w(z)$ in our method
is equivalent to dropping ``small eigenvalues'' in a
principal component analysis. Both methods have in common the 
possibility to obtain the shape of the function $w(z)$ and are
expected to provide good reconstructions with large samples. 
The approach explored here enables to see the filter that
 we are placing between the true model and the 
estimated one, when asking questions about $w(z)$ given a certain sample.

The exploration of $w(z)$ applying this method to the present sample 
helps to answer the question on whether there is evidence in 
the evolution of $w(z)$ along z. 
We find that the highest degree of information on $w(z)$ from 
present samples is at z $\sim$ 0.2--0.5. The current SNe Ia data indicate a
tendency towards $w=0$ at those intermediate redshifts. However, 
though this feature is consistently found in the analysis when adding
other cosmic distance indicators, the
cosmological constant is still within the 2$\sigma$ level contours of
the solution of $w(z)$. Moreover, it is found that there is
no information on $w(z)$ at z $>$ 0.6 to imply any possible
evolution of $w(z)$ at high z.  Adding the data set of FR IIb
radiogalaxies to SNe Ia enhances the significance level of
 the peak towards $w(z) = 0$ in the
continuos reconstruction as well as in a discrete one which shows the 
present value of $w(z)$  ($w_{0}$)
and its first derivative ($w_{a}$).

Retrieving $w(z)$ will improve when using a 
large sample of SNeIa data contributed by SNAP and 
other cosmological probes at z $>$ 1. This approach helps to explore the
possibility of recovering a wide variety of dark energy models, and
differentiate them from the cosmological constant.

\begin{acknowledgments}
This work has been partially supported by the European Research and
Training Network Grant on Type Ia Supernovae (HPRN--CT--20002-00303),
and by research grants in cosmology by the Spanish DGYCIT
(ESP20014642--E and BES-2004-4435) and Generalitat de Catalunya 
(UNI/2120/2002). 
\end{acknowledgments}

\end{document}